\documentclass[traditabstract]{aachanged}
\usepackage[english]{babel} 
\usepackage{amsmath}
\usepackage{amssymb}
\usepackage{hyperref}
\usepackage{graphicx}
\renewcommand{\tablename}{table}

\newcommand{\qso}{QSO$\,$1232+082}

\begin{document}

\title{Directional Radiation and Photodissociation Regions \\
in Molecular Hydrogen Clouds}

\author{S. A. Balashev\thanks{balashev@astro.ioffe.ru} \and D. A. Varshalovich\thanks{varsh@astro.ioffe.ru} \and A. V. Ivanchik\thanks{iav@astro.ioffe.ru}}

\date{}

\institute{
{\qquad \qquad \qquad \quad \it{Ioffe Physical-–Technical Institute, Politekhnicheskaya 26, St. Petersburg, 194021 Russia}}}


\abstract{
Some astrophysical observations of molecular hydrogen point to a broadening of the velocity
distribution for molecules at excited rotational levels. This effect is observed in both Galactic and high redshift
clouds. Analysis of H$_2$, HD, and C\,I absorption lines has revealed the broadening effect in the
absorption system of \qso \,\,($z_{\rm abs}=2.33771$). We analyze line broadening mechanisms by
considering in detail the transfer of ultraviolet radiation (in the resonance lines of the Lyman and Werner H$_2$
molecular bands) for various velocity distributions at excited rotational levels. The mechanism we suggest
includes the saturation of the lines that populate excited rotational levels (radiative pumping) and manifests
itself most clearly in the case of directional radiation in the medium. Based on the calculated structure of
a molecular hydrogen cloud in rotational level populations, we have considered an additional mechanism
that takes into account the presence of a photodissociation region. Note that disregarding the broadening
effects we investigated can lead to a significant systematic error when the data are processed.
}


\keywords{high-redshift molecular clouds - absorption lines in quasar spectra - radiative transfer - \qso.}


\maketitle

\section*{INTRODUCTION}
\noindent
Molecular hydrogen, the most abundant molecule
in the interstellar medium, is observed not only in
our Galaxy (\cite{Savage1977}; \cite{Shull1982}; \cite{Rachford2002}) and Local Group galaxies
(\cite{Tumlinson2002}; \cite{Bluhm2003}) but also in
the spectra of high-redshift quasars (see \cite{Levshakov1985}; \cite{Noterdaeme2008}).


The importance of molecular hydrogen, particularly
at the early evolutionary phases of the Universe,
lies in the fact that it plays a key role in the theory of
star formation, being the main coolant. Observations
of various rotational H$_2$ states allow the main physical
parameters (kinetic temperature, pressure, particle
density, etc.) to be determined in the molecular hydrogen
clouds that existed 12 Gyr ago. The relative population
of rotational H$_2$ levels also makes it possible to
estimate the intensity of ultraviolet radiation, whose
high value can serve as an indicator of active star
formation in early protogalaxies. In addition, precision
measurements of molecular hydrogen wavelengths
at high redshifts are used to estimate the possible
cosmological change in the proton-to-electron mass
ratio (see, e.g., \cite{Ivanchik2005}).

\begin{figure*}
	\centering
		\includegraphics[width=0.70\textwidth]{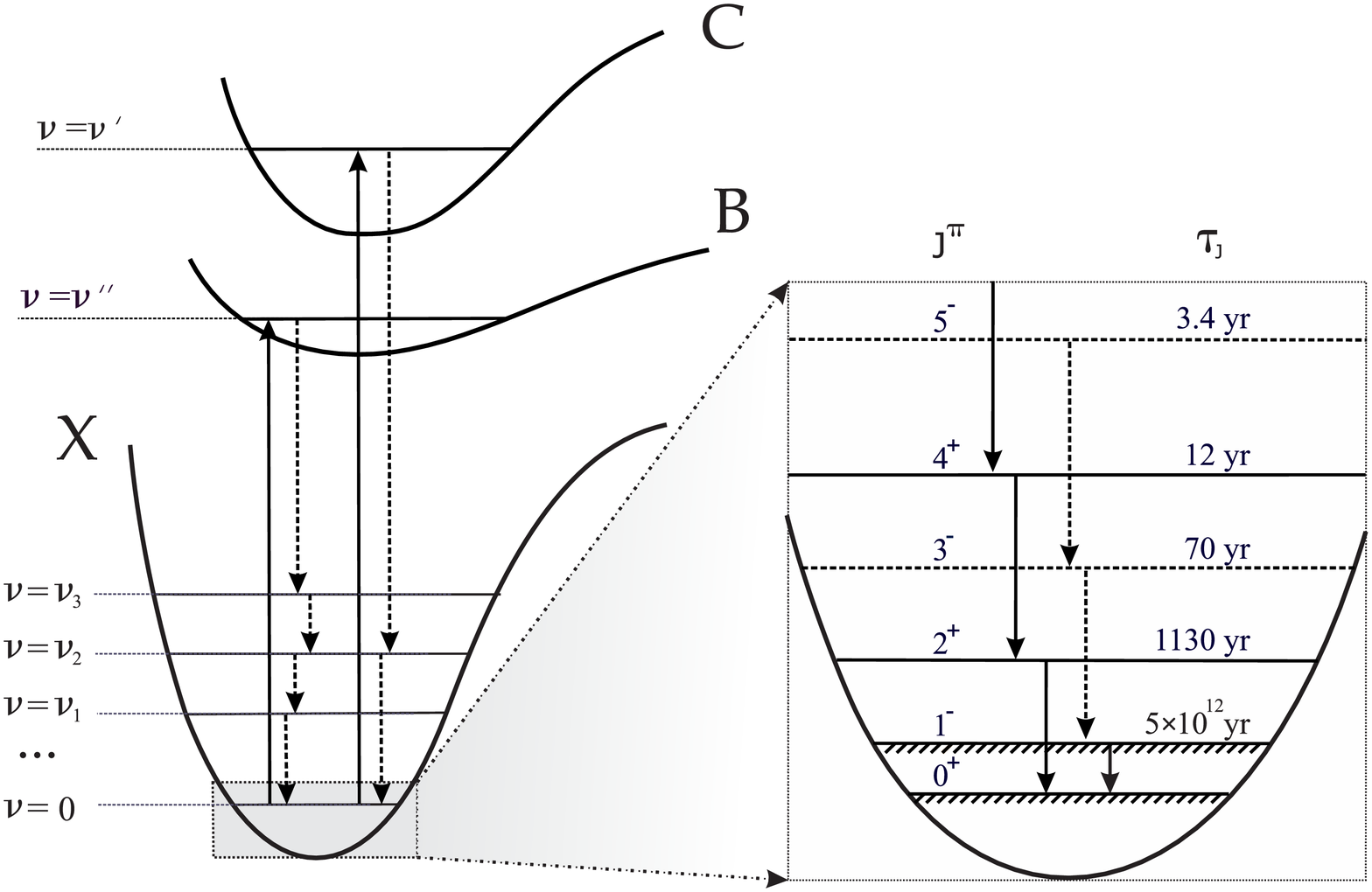}
		\caption{Level diagram for the hydrogen molecule. The ground X$ ^1\Sigma_g^+$ and excited B$ ^1\Sigma_u^+$  and C$ ^1 \Pi_u$ electronic states are presented. The vibrational structure of the electronic levels is shown. The rotational structure for the lower vibrational, ground electronic state, in which the times of spontaneous transitions are indicated, is presented on the right panel.}
		\label{H2_electron}
\end{figure*}

The ground X$ ^1\Sigma_g^+$ and excited B$ ^1\Sigma_u^+$ and C$ ^1 \Pi_u$
electronic states are distinguished in the molecular
hydrogen level structure. There are vibrational and
finer rotational level structures in each electronic state
(see Fig. \ref{H2_electron}).


Since the dipole transitions are forbidden in the
ground electronic state, molecular hydrogen is commonly
observed using the system of lines between the
ground X state and the excited B and C states, which
are called the Lyman and Werner bands, respectively.
These bands lie in the ultraviolet spectral range.
However, for hydrogen at high redshifts (z$\,\,\sim\,\, 2 \div 4$),
the absorption lines fall within the visible spectral
range. Therefore, it can be observed using groundbased
telescopes with a high spectral resolution.


Another peculiarity of the hydrogen molecule is
the mechanism of its dissociation. The point is that
the dissociation continuum of molecular hydrogen is
higher than that of atomic one. This means that there
is no direct photodissociation of molecular hydrogen
in the interstellar medium, because atomic hydrogen
depletes the spectrum above 13.6 eV. The dissociation
proceeds through the mechanism that was first
described by \cite{Stecher1967}. After their
excitation in the Lyman or Werner bands, about 87\,\%
of the molecules return rapidly, in a time \mbox{$\tau \sim 10^{-6}\, \mbox{s}$},
to the ground electronic state at upper ro-vibrational
levels, while 13\,\% of the molecules fall into the continuum,
i.e., dissociate (\cite{Abgrall1992}). The undissociated
molecules fall into the lower vibrational state
through cascade transitions and populate the long-lived
rotational levels (see \cite{Dalgarno1974}).
This mechanism is called radiative pumping.


The absence of dipole transitions in the lower vibrational,
ground electronic state not only makes
the rotational levels basically metastable but also
strongly suppresses the direct radiative population.
The typical physical conditions in molecular hydrogen
clouds are \mbox{$T\sim 50 \div 150\,\mbox{K}$}, \mbox{$n \sim 10 \div 1000 \, \mbox{sm}^{-3}$} 
(\cite{Rachford2002}, \cite{Shull2000}). At such
parameters, the collision rate is too low (compared to
the rates of spontaneous transitions) for the excited
rotational levels to be populated. Therefore, being in
background ultraviolet radiation with an intensity 
$I \sim  10^{-18}\, \mbox{erg/sm}^2\mbox{/s/Hz}$ (for our Galaxy, at wavelengths
of the order of $\lambda = 1000\,$\AA, \cite{Draine1978}), the
excited rotational $J>2$ levels are populated mainly
through radiative pumping.


Fourteen absorption systems of molecular hydrogen
have been identified to date in the spectra of
distant quasars (\cite{Noterdaeme2008}, \cite{Srianand2008}). 
One of these belongs to the spectrum
of \qso. The peculiarity of this system
is that the effective width parameter of the velocity
distribution $b$ for the excited rotational levels of hydrogen
molecules is larger than the parameter $b$ for HD
molecules and C\,I atoms (Ivanchik et al., to be published).
In this paper, we will show how the propagation
of directional radiation can broaden the velocity
distribution of H$_2$ molecules at the levels populated
by radiative pumping. In addition to our effect, we will
consider a model that takes into account the presence
of a photodissociation region, as was suggested by
\cite{Noterdaeme2007}.


\section*{OBSERVATIONAL DATA}
\label{QSO1232}
\noindent

\subsection*{The Absorption System in the Spectrum \qso}

The absorption system in the spectrum of
\qso\,\, has been known since 1999 (\cite{Ge1999}). 
It was measured in more detail by
\cite{Petitjean2000} with the 8.2-m VLT telescope
using theUVES echelle spectrograph. Subsequently,
in 2001, HD molecules were first detected in this
system at high redshifts (\cite{Varshalovich2001}).
This had long been the only identification of HD
molecules at high redshifts (the second system was
identified by \cite{Srianand2008}). Analysis of this
system revealed evidence for a broadening of the
velocity distribution of molecules at excited rotational
levels of molecular hydrogen. This effect is important
in that assigning the same Doppler parameter to all
rotational levels leads to a systematic error in the
column density and, as a result, in other physical
parameters of the system.


\begin{figure*} 
	\centering
		\includegraphics[width=0.7\textwidth]{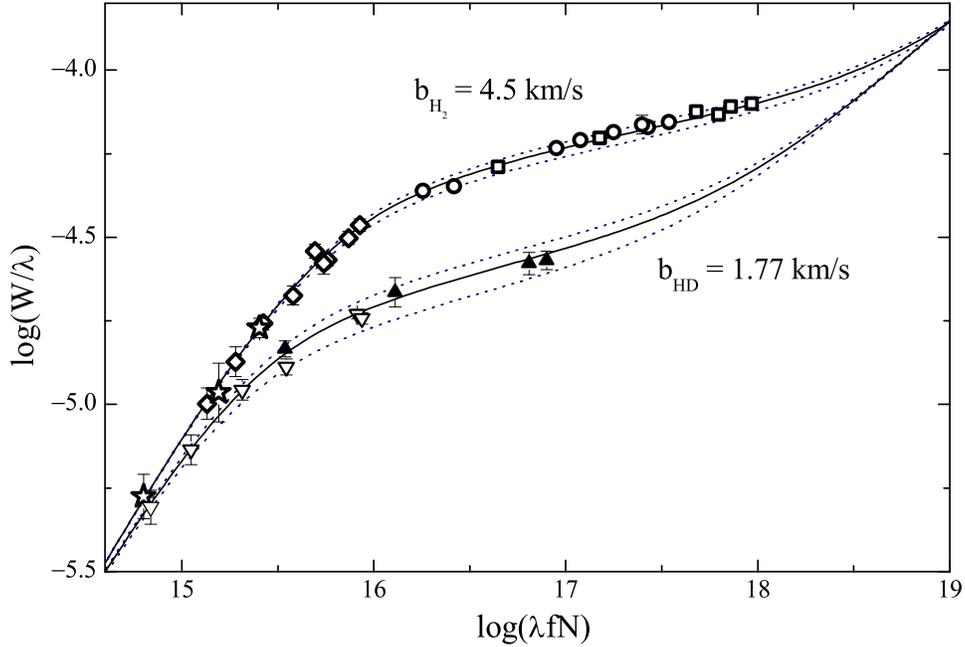} 
		\caption{Standard curves of growth for the H$_2$, HD, and C\,I lines corresponding to the z = 2.33771 absorption system of
\qso. The equivalent widths for various excited rotational H$_2$ levels are plotted: $J=2$ (squares), $J=3$ (circles),
$J=4$(diamonds), and $J=5$(asterisks). The equivalent widths of the transitions from the excited rotational levels fall best on
the standard curve of growth with $b=4.5 \, \mbox{km/s}$. Also shown are the HD (upward triangles) and CI (downward triangles)
data. The standard curves of growth with $b=1.77 \, \mbox{km/s}$ и $b=1.67 \, \mbox{km/s}$ are presented for HD and C\,I, respectively.}
		\label{QSOdata}
\end{figure*}


The data on the absorption system of
\qso\,\, used here are presented in Fig. \ref{QSOdata}.
We used the curve-of-growth method to analyze the
data. The curve of growth shows the dependence of
the line equivalent width on the column density. Since
it also depends on the transition line parameters, the
column density of an element and the distribution
parameters can be determined in certain cases given
the set of various lines. For a Maxwellian distribution
(with the Doppler parameter $b$), the curve of growth
is said to be a standard one. In our paper, we use a
nonequilibrium velocity distribution to construct the
curve of growth and call this curve a “nonstandard”
one. Figure \ref{QSOdata} shows the equivalent widths for various
excited rotational levels of H$_2$ ($J=2,3,4,5$), HD,
and C\,I. In addition, the standard curves of growth are
plotted in the figure for $b=4.5 \, \mbox{km/s}$ и $b=1.77 \, \mbox{km/s}$. The
equivalent widths for the HD lines (upward triangles)
fall best on the curve of growth with $b=1.77 \, \mbox{km/s}$.
The data on carbon (downward triangles) for this
cloud have the Doppler parameter $b=1.67 \, \mbox{km/s}$. It
thus follows that the equivalent widths of the H$_2$ lines
should have had $b$ equal to $b=1.82 \, \mbox{km/s}$. However,
the excited rotational levels of molecular hydrogen fall
best on the curve of growth with $b=4.5 \, \mbox{km/s}$, i.e.,
they have a distribution that is two and a half times
broader than that expected from our analysis of the
HD and C\,I data. Note that the Doppler parameter $b$
for the $J=0, 1$ rotational levels cannot be determined
by the curve-of-growth method, because the absorption
lines are strongly saturated and their equivalent
widths fall on the square-root segment where the
curves of growth with various $b$ merge together. Thus,
a broadening of the velocity distribution at excited
rotational levels of molecular hydrogen manifests
itself in this system.

\subsection*{Observations of the Broadening Effect in Other Systems}

A broadening of the effective Doppler parameter $b$
with increasing rotational level $J$ was observed previously
for our Galaxy (\cite{Spitzer1973}, \cite{Jenkins1997}, \cite{Lacour2005}) and
in a high-redshift system \cite{Noterdaeme2007}). 
\cite{Spitzer1973} were the first to notice
this effect while analyzing the curve of growth for observations
with the ultraviolet telescope of the Copernicus
Orbital Observatory. However, since the data
had large errors, it is hard to say whether this effect is
systematic. Much later, the observations by 
\cite{Jenkins1997} on IMAPS revealed a broadening
in molecular hydrogen toward \mbox{$\zeta \,$ Ori A}. Several
years ago, while observing excited rotational levels of
molecular hydrogen in the directions of four young
stars from the FUSE satellite, \cite{Lacour2005}
found that such an effect was present in all four
directions. The fourth and most recent observation
refers to extragalactic molecular hydrogen. 
\cite{Noterdaeme2007} revealed this effect for the absorption
system at redshift $z=2.402$ in the spectrum of
the quasar HE 0027–-1836.


Note that no satisfactory explanation for the
broadening effect has existed until now. Different
authors use different models to explain it. The system
observed by \cite{Jenkins1997} is peculiar
in that the line profiles exhibit three components
for which the column densities at rotational levels
are low ($N_J \lesssim 10^{15}\, \mbox{sm}^{-2}$ and that there exists
a systematic shift in the velocities with increasing
rotational level. This enables the authors to explain
the effect in question by a model with a shock. The
observed broadening and the velocity shifts stem
from the fact that the flow from the shock front
gradually cools down and decelerates. This model
runs into difficulties in explaining the data in which
one component of molecular hydrogen is observed,
there are no velocity shifts for various rotational levels,
and the H$_2$ column densities are high.


\cite{Lacour2005} assume that during its formation
on dust, the hydrogen molecule acquires an
additional energy of $\sim 4.5\, \mbox{eV}$ that is distributed between
the vibrational, rotational, and translational
degrees of freedom. In other words, the molecules are
formed in excited states with an excess kinetic energy.
However, the authors point out that the formation rate
of molecules on dust should be an order of magnitude
higher to explain the broadening effect. This is
particularly problematic for the high-redshift absorption
systems of molecular hydrogen associated with
damped Ly-$\alpha$ systems; observations of the metallicity
in the latter (\cite{Ledoux2003}, \cite{Noterdaeme2008}) 
show that the dust content in them is more than
an order of magnitude lower than that in our Galaxy.
The damped Ly-$\alpha$ systems are the systems with very
high atomic hydrogen column densities (N(H\,I) $> 10^{20}\, \mbox{sm}^{-2}$)
 that are observed in quasar spectra and
that are probably high-redshift protogalaxies.


Another explanation for the broadening effect was
also offered by \cite{Noterdaeme2007}. They showed
that the broadening could be explained by the presence
of two components in the cloud. The first component
is under the influence of strong background
ultraviolet radiation, which increases the radiative
pumping of excited rotational levels and heats up the
medium through hydrogen dissociation—this is the
so-called photodissociation region (\cite{Hollenbach1999}). 
In the second component, the ultraviolet
radiation is appreciably weaker; therefore, it is
colder and the excited levels are populated sparsely.
In that case, the excited and ground rotational levels
manifest themselves mainly in the hot and cold components,
respectively, which can provide the observed
effect. As will be shown below, when the radiative 
transfer in the Lyman and Werner bands is considered,
these regions naturally emerge in the same
molecular cloud.


Note that \cite{Lacour2005}, \cite{Noterdaeme2007}
 assumed that radiative pumping could
not produce a broadening. The authors placed emphasis
on the fact that the molecule did not change
its velocity during radiative pumping transitions and
sought for other possible explanations. This assertion
is beyond doubt. However, we will show in the next
section that when the transfer of directional radiation
is considered in detail, a broadening at the levels being
populated can arise during radiative pumping if the
population from the line wings is taken into account.

 
\section*{MAIN POINTS OF THE MODEL}
\label{2levelsys}
\noindent

\begin{figure}
	\centering
		\includegraphics[width=0.40\textwidth]{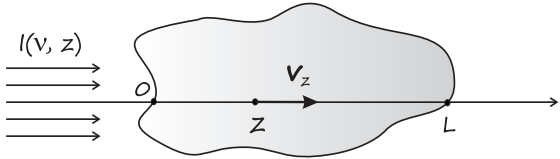} 
		\caption{Geometric model of cloud. Directional radiation with
an intensity $I(\nu,z)$ falls on the cloud.} 
		\label{geom_model}
\end{figure}

A broadening of the velocity distribution at the
levels populated by radiative pumping can be obtained
by considering the radiative transfer. The broadening
effect will be at a maximum during the propagation
of directional radiation. Therefore, let us consider a
molecular hydrogen cloud on which directional radiation
with an intensity $I(\nu, z)$ falls (see Fig. \ref{geom_model}).
We will be interested in the velocity distribution of
the molecules populated by radiative pumping as a
function of the depth of radiation penetration into the
cloud.


\begin{figure}
	\centering
		\resizebox{!}{0.2\textwidth}{\includegraphics{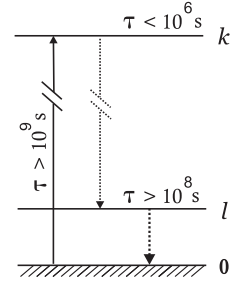}}
		\caption{Three-level system: $0$ is the ground level from
which level $l$ is radiatively pumped through the resonance
level $k$. The characteristic time scales of the processes
considered in this system corresponding to a hydrogen
molecule in the interstellar medium are shown near the
arrows indicating these processes.} 
		\label{elect2}
\end{figure}

To understand the effect qualitatively, let us consider
a three-level system (see Fig. \ref{elect2}). Naturally, this
system will fully reflect the main physical properties
of the hydrogen molecule of interest to us and it can
be used as an example to obtain the broadening effect
analytically. Thus, we have the ground level ($0$) that
acts as a reservoir. A metastable level $l$ is radiatively
pumped from the ground level through the system’s
photoexcitations to level $k$ followed by its relaxation
to level $l$. The direct transitions from the ground level
to level $l$ will be assumed to be forbidden.


Based on the characteristic time scales shown
in Fig. \ref{elect2} (which correspond to those for the hydrogen
molecule), we can assume that the system
passes from level $k$ to level $l$ almost instantly. Since
level $l$ is metastable, we can observe the molecules
at this level. To obtain the velocity distribution of
the molecules at level $l$, let us consider the balance
equation for the molecules at the level

\begin{equation}
	 	n_l(v_z, z)  A_l =n_0(v_z, z) P_{0k}(v_z, z)
	 	\label{balance}
\end{equation}
The left-hand and right-hand sides of the equation
describe, respectively, the spontaneous downward
transitions and the population of level $l$ from the
ground level through radiative pumping via level $k$.
$P_{0k}$ is the capability of the particles to pass to level $k$
under radiation or the fraction of the particles that
pass to level $k$ per unit time at depth $z$ and that have a
velocity $v_z$; in other words, $P_{0k}$ is the photoexcitation
rate:

\begin{equation}
   	P_{0k}(v_z, z)= \int\limits_{-\infty}^{+\infty}\frac{I(\nu,z)}{h\nu} \kappa_{0k}\left(\nu+ \nu \frac{v_z}{c}\right) d\nu.
   	\label{P_0k}
\end{equation}
It is important to note that this quantity is a function
of $(v_z, z)$.

We can neglect the multiple scatterings while
considering the radiative transfer in the cloud, since
the resonance photons that accomplish radiative
pumping fragment as the excited states decay due
to the presence of a ro-vibrational cascade in the H$_2$
molecule. Therefore, we can use a radiative transfer
equation in the form

\begin{equation}
 		I(\nu,z)= I(\nu, 0) e^{-\tau_{0k}(\nu, z)},
 		\label{intens}
\end{equation}
where
\begin{equation}
 		\tau_{0k}(\nu, z)=\int\limits_0^zdz'\int\limits_{-\infty}^{+\infty}dv_z n_0(v_z, z')\kappa_{0k}\left(\nu+ \nu \frac{v_z}{c}\right).
 		\label{tau}
\end{equation}
The quantity $\kappa_{0k}$ in Eqs. \eqref{P_0k} and \eqref{tau} is the photoexcitation
cross section for the $0 \to k$ transition.

\subsection*{The Broadening Effect}

\begin{figure*}[ht!]
	\centering
		\includegraphics[width=0.8\textwidth]{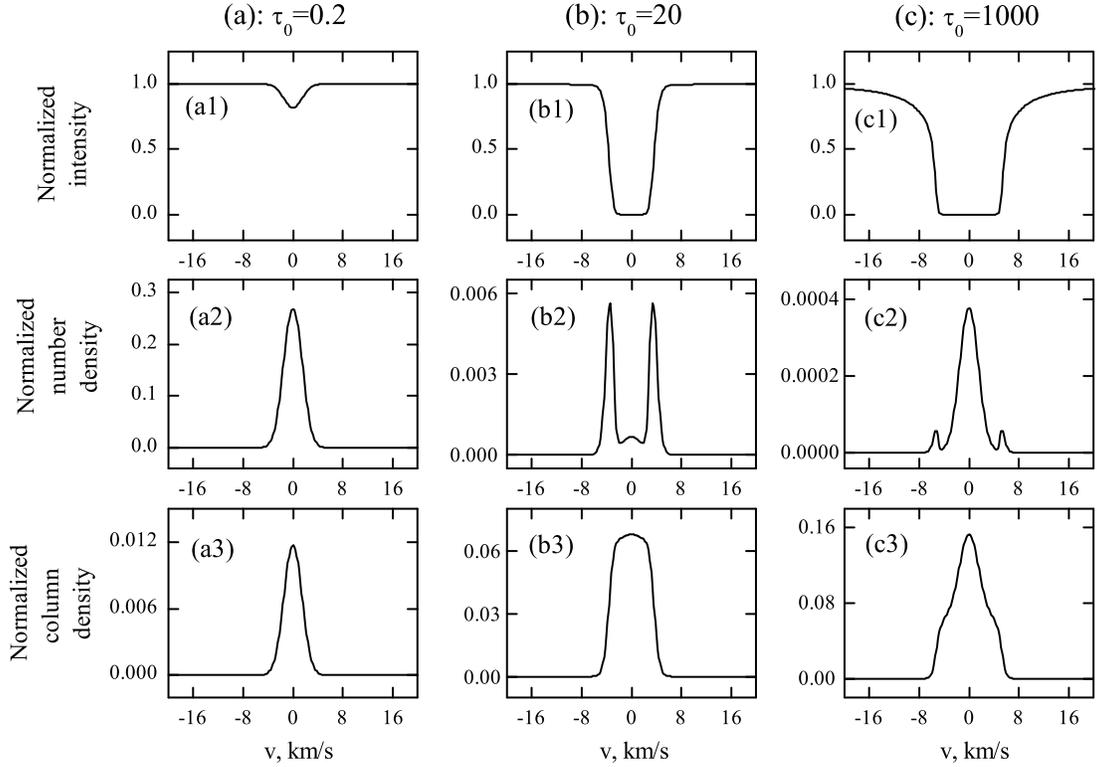}
		\caption{Spectral resonance line profiles and velocity distributions of molecules. The calculation is presented for three optical
depths at the line center: (a) $\tau_0= 0.2$, (b) $\tau_0= 20$, and (c) $\tau_0= 1000$. The upper panels show the profile of the line through which level $l$ is radiatively populated. The middle panels show the normalized molecular number density distribution produced
by radiative pumping. The lower panels show the normalized molecular column density distribution integrated along the line
of sight to the depth corresponding to the line profile on the upper panels.}
		\label{populating}
\end{figure*}

An expression for the velocity distribution of the
molecules at level $l$ can be obtained by rewriting
Eq. \eqref{balance} using Eq. \eqref{P_0k}:
\begin{equation}
		n_l(v_z, z) =\frac{n_0(v_z, z)}{A_l} \int\limits_{-\infty}^{+\infty}\frac{I(\nu,z)}{h\nu} \kappa_{0k}\left(\nu+ \nu \frac{v_z}{c}\right) d\nu.
	 \label{n_l}	 
\end{equation}
The distribution at level $l$ is broadened during the
formation of an absorption line in the $0 \to k$ transition
followed by the population of level $l$ (Fig. \ref{elect2}). Figure \ref{populating}
presents the calculated number and column densities
and line profiles for typical molecular hydrogen line
parameters (we chose the L3-0R(0) line). In what
follows, we assume that the distribution at the ground
level $0$ is Maxwellian.


The upper panels in Fig. \ref{populating} (a1, b1, c1) present
the line profile for a given optical depth $\tau_0$ at the
line center (indicated above the panels). The middle
panels (a2, b2, c2) show the normalized distribution
of the number density populated by radiative pumping
to level $l$ at depth $z$ where the line profile was formed.
The presented number density distributions were
normalized to the area under the distribution for
$\tau_0=0.2$ (panel (a2)). Therefore, the area under the
distributions for $\tau_0=20$ (panel (b2)) and $\tau_0=1000$
(panel (c2)) shows the extent to which the total
photoexcitation rate decreases compared to the depth
where $\tau_0=0.2$. The lower panels (a3, b3, c3) display
the normalized distribution of the column density,
i.e., integrated over the line of sight to depth $z$. The
column density distributions were normalized to the
area of the distribution under its curve for $\tau_0=1000$ 
(panel (c3)). The velocity in $\mbox{km/s}$ is along the
 $x$ axes (for the line profile, this corresponds to the
Doppler wavelength shift). We see that as long as
the line profile is unsaturated ($\tau_0=0.2$, (a1)), the
populated (a2) and column (a3) densities correspond
to a Maxwellian distribution at the ground level, i.e.,
no broadening occurs. When the lines are saturated
($\tau_0=20$, (b1)), the distribution populated to level $l$
becomes a highly nonequilibrium one. As long as the
absorption is a Doppler one, the molecule absorbs
a photon that closely corresponds to its Doppler
velocity shift. In other words, the photoexcitation
cross section $\kappa_{ok}$ in Eq. \eqref{n_l} can be replaced with
a delta function; therefore, the dependence of the
photoexcitation cross section $P_{0k}$  on the velocity of
the molecule will closely correspond to the profile
of the $0 \to k$ transition line. When the line is saturated,
the photons are deficient at the line center
and the population takes place mainly at the line
edges, where the photons are plenty. Therefore, the
number density distribution at level $l$ becomes a
highly nonequilibrium one (panel (b2)). The column
density distribution (panel (b3)) is an observable
quantity in the spectra; it exhibits a broadening with
the formation of a plateau.

As the line is saturated further ($\tau_0=1000$, (c1)),
i.e., as the radiation penetrates deeper into the
medium, the Lorentz wings will manifest themselves
in the line profile. This means that level $l$ will also
be populated in the Lorentz wings. In this case, the
transition wavelength does not corresponds to the
Doppler velocity shift and the replacement of  $\kappa_{0k}$ 
with a delta function is invalid. This population is
peculiar in that the distribution populated by the
Lorentz wings will correspond to the distribution at
the ground level, i.e., it will be Gaussian. We see from
panel (c2) that even a small fraction at a given optical
depth corresponds to a nonequilibrium distribution.
Thus, the population in the Lorentz wings can greatly
reduce the broadening of the column density distribution
(panel (c3)).
	 
 
 The nonequilibrium distribution populated to level $l$
will manifest itself in the absorption lines from this
level, while the lines will not be described by a
Voigt profile. When a Voigt profile is fitted into the
line, the effective width parameter of the Maxwellian
distribution $b$ will naturally be larger than that for
the ground level. Therefore, when we analyze the
absorption lines for level $l$ and use the standard theory
of radiative transfer, the parameter $b$ to be determined
will effectively increase.

 
 \subsection*{Thermalization and Turbulence}
 	
 	The nonequilibrium distribution populated to level $l$
can be thermalized with time or remain broadened in
the presence of turbulent motions. It all depends on
the relationships between the rates of characteristic
processes. To describe the turbulent motions, we will
use the assumption about microturbulence under
which the distribution in the cloud in turbulent
velocity $w_{turb}$ is described by a Gaussian function with
a parameter $b_{turb}$. The ground level may be considered
to be thermalized, i.e., the velocity distribution of the
molecules related to thermal motion, $w_{th}$, is described
by a Maxwellian distribution with a parameter $b_{th}$.
The total distribution obtained by a convolution of
the turbulent and thermal distributions will then
be Maxwellian with the parameter $b=\sqrt{b^2_{turb}+b^2_{th}}$
Therefore, the relationship between the following
three rates of the processes should be considered: the
thermalization rate, the turbulization rate, and the
photoexcitation rate from level $l$. The turbulization
rate may be considered to be much lower than the
other two rates. If the thermalization rate is higher
than the photoexcitation rate from level $l$, then the
distribution that manifests itself in the absorption line
from level $l$ will be thermalized.

  
  The distribution thermalization effect for static
turbulence can be written as
\begin{equation}
	 \tilde n_l(v_z, z)= \int\limits_{-\infty}^{+\infty}dv'_z \, n_l(v'_z, z)  w_{+}\left(v_z-v'_z\frac{b^2_{turb}}{b^2}\right).
	 \label{thermalize}
\end{equation}
(for the derivation, see the Appendix).
	
	When $b_{th} \to 0$ \,  $w_+(v) \to \delta(v)$ and, hence,
$\tilde n_l(v,z)\equiv n_l(v,z)$, i.e., the nonequilibrium distribution
is retained. When $b_{turb}\to 0$, $w_{+}$ does not depend
on $v_z$ and the velocity distribution is completely
thermalized, acquiring a Gaussian shape with $b=b_{th}$.


\subsection*{Optical Depths of Broadened Lines}

As a result, we find that, given thermalization \eqref{thermalize},
the distribution at level $l$ will be
\begin{multline}
		 \tilde n_l(v_z, z)=\frac{n_0(z)}{A_l} \int\limits_{-\infty}^{+\infty}dv'_z \, P_{0k}(v'_z, z) \times \\
	 \times n_0(v'_z,z)   w_{+}\left(v_z-v'_z\frac{b^2_{turb}}{b^2}\right).
   \label{n_l_over}
\end{multline}

As we showed above, the particle velocity distribution
at level $l$ \eqref{n_l} can manifest itself in the absorption
line associated with the transition from this level to
some resonance level $m$. We will assume that the
absorption line profile is defined by the optical depth
in accordance with Eq. \eqref{intens}. The optical depth $\tau_{lm} \,$ for
the $l\to m \quad$ transition will have the form \eqref{tau} with the
cross section $\kappa_{lm}$ corresponding to the transition and
the number density $\tilde n_l(v_z, z)$ given in \eqref{thermalize}. Thus, we
obtain
\begin{multline}
   \tau_{lm}(\nu, L)=\int\limits_0^Ldz\int\limits_{-\infty}^{+\infty}dv_z   \kappa_{lm}\left(\nu+\nu \frac{v_z}{c}\right) \tilde n_l(v_z,z) = \\   =\int\limits_0^Ldz\int\limits_{-\infty}^{+\infty}dv_z   \kappa_{lm}\left(\nu+\nu \frac{v_z}{c}\right) \times \\ \times\int\limits_{-\infty}^{+\infty}dv'_z  n_l(v'_z, z)  w_{+}\left(v_z-v'_z\frac{b^2_{turb}}{b^2}\right). 
   \label{taulm}
\end{multline}

It is important to note that in our approximations,
the $z$ dependence is present only in $n_l(v_z, z)$. Using
Eq. \eqref{n_l}, we can represent the expression for the optical
depth as
\begin{equation}
   \tau_{lm}(\nu)=\int\limits_{-\infty}^{+\infty}dv'_z \, W_l(v'_z)  \tau^+_{lm}\left(\nu+\nu_{lm}\frac{v'_z}{c}\frac{b^2_{turb}}{b^2}\right),
   \label{taulm2}
\end{equation}
where $W_l(v_z)$  is the nonequilibrium distribution of the
particles thrown to level $l$ integrated along the line of
sight:
\begin{equation}
	  W_l(v_z)=\frac{N_l(v_z)}{\int\limits_{-\infty}^{+\infty}dv_z \, N_l(v_z)}=\frac{\int\limits_0^Ldz\, n_l(v_z, z)}{\int\limits_0^Ldz \int\limits_{-\infty}^{+\infty}dv_z\, n_l(v_z, z)}.
\end{equation}

The quantity $\tau^+_{lm}$ in Eq. \eqref{taulm2} is the optical depth in
the standard theory of radiative transfer, which can be
written via the Voigt function as 
\begin{equation}
	 \tau^+_{lm}\left(\nu+\nu_{lm}\frac{v'_z}{c}\frac{b^2_{turb}}{b^2}\right)=\sqrt{\pi} \frac{e^2}{m_ec}   \frac{4\pi N_l f_{lm}  a_{lm}}{\gamma}  H(a_{lm}, x),
\end{equation}
where
\begin{align*} 
   &a_{lm}=\frac{\gamma c}{4\pi\nu_{lm}b_{th}\sqrt{1+(b_{turb}/b)^2}}, \\
   &x=-\frac{c}{b_{th}\sqrt{1+(b_{turb}/b)^2}} \left(1+\frac{v'_z}{c}\frac{b^2_{turb}}{b^2}-\frac{\nu_{lm}}{\nu}\right).
\end{align*}
We can write expressions for $N_l(v'_z)$ — the column
density distribution of the particles that are populated
to level $l$ and that have no time to be thermalized.
Using Eq. \eqref{n_l}, we will find that 
\begin{equation}
	 N_l(v_z)=\int\limits^L_0 dz\, n_l(v_z, z)= \frac1{A_l} \int\limits^L_0 dz\, n_0(v_z, z)  P_{0k}(v_z, z).
\end{equation}

For a uniform and equilibrium distribution at the
ground level and for the radiative transfer equation
written for the $i\to k$ lines in form \eqref{intens}, we can easily
obtain
\begin{multline}
	 N_l(v_z)= \frac{N_0}{A_l}  w(v_z)  \int\limits_{-\infty}^{+\infty}d\nu\, \frac{I_0(\nu)}{h\nu} \times \\
	 \times \frac{(1-e^{-\tau_{0k}(\nu, L)})}{\tau_{0k}(\nu, L)}  \kappa_{0k}\left(\nu+ \nu \frac{v_z}{c}\right),
	\label{N_l}
\end{multline}
where $I_0(\nu)=I(\nu, z=0)$.

Passing to the limit in Eq. \eqref{taulm2} allows physically
meaningful expressions to be derived for the extreme
cases:
\begin{enumerate}
	\item The absence of turbulence ($b_{turb} \ll b_{th}$).
	
	In this case, $b_{th} \approx b$. Then,
	\begin{equation}
      	\tau_{lm}(\nu)= \int\limits_{-\infty}^{+\infty}dv'_z   W_l(v'_z)   \tau^+_{lm}\left(\nu\right)= \tau^+_{lm}(\nu),
  \end{equation}
  		where $\tau^+_{lm}$ is the optical depth in the standard theory of
radiative transfer with a thermal broadening $b=b_{th}$.
Thus, in this case, the nonequilibrium distribution radiatively
populated upward is completely thermalized
to become an equilibrium one and gives a standard
shape for the line profile.
	
	\item Dominant turbulence ($b_{turb} \gg b_{th}$).
	 		
		In this case, $b_{turb} \approx b$ and then $\tau^+_{lm}\left(\nu+\nu_{lm}\frac{v_z}{c}\frac{b^2_{turb}}{b^2}\right)$
will correspond to the cross section
$\kappa_{lm}\left(\nu+\nu \frac{v_z}{c}\right)$ to constant coefficients and the
broadening effect related to radiative pumping will
manifest itself in full:  	
	   \begin{equation}
	      \tau_{lm}(\nu, L)=\int\limits_{-\infty}^{+\infty}dv_z    \kappa_{lm}\left(\nu+\nu \frac{v_z}{c}\right) N_l(v_z).
     \end{equation}

\end{enumerate}

\subsection*{“Nonstandard” Curves of Growth}

\begin{figure*}[ht!]
	\centering
		\includegraphics[width=0.7\textwidth]{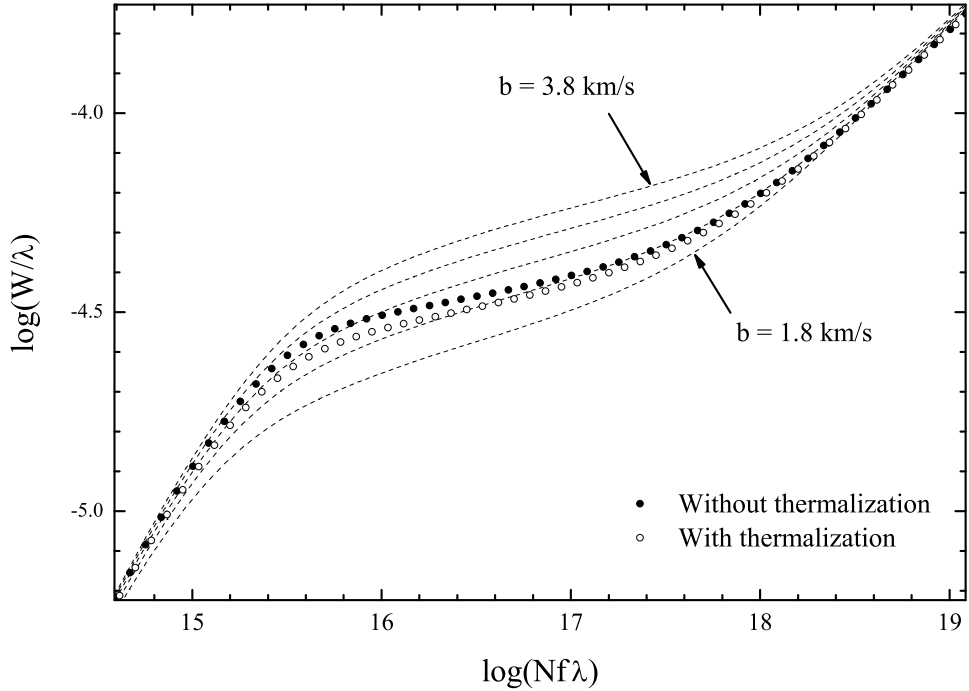}
		\caption{Nonstandard curves of growth. The circles indicate the curves of growth for the velocity distribution of the molecules
broadened by radiative pumping: the filled and open circles indicate the results of our calculation, respectively, without and with
thermalization of the velocity distribution at the level. The following distribution parameters were chosen: $b_{th} = 0.8\,\mbox{km/s}$, $b_{turb} = 1.6\,\mbox{km/s}$. The dashed lines represent the standard curves of growth with parameters $b$ from 1.8 to 3.8 km/s at steps of 0.5 km/s. In the absence of a broadening effect, the nonstandard curves of growth marked by the circles would coincide with the lower standard curve of growth with $b = 1.8\,\mbox{km/s}$.}
		\label{nonstandart_cog}
\end{figure*}

Using the above formulas, we can demonstrate
how the broadening effect will appear in curve of growth
representation. In Fig. \ref{nonstandart_cog}, the dashed lines
indicate the standard curves of growth for distribution
parameters $b$ from 1.8 to 3.8 km/s at steps of
0.5 km/s. In the same figure, the circles indicate the
curves of growth for the broadening effect that we calculated
based on the three-level model. The curves of
growth correspond to the transitions from level $l$, i.e.,
to the particle velocity distribution at level $l$. Since, as
was shown above, this distribution can differ from the
Maxwellian one, the curves of growth will also differ
from the standard ones. The figure shows two limits:
with (open circles, see \eqref{thermalize}) and without (filled circles)
thermalization.


The shape of the nonstandard curve of growth
depends on certain parameters of the calculation.
First, these include the parameter $b$, the width of
the Maxwellian distribution at the ground level from
which the population occurs. For the calculation in
question, btot is $b_{tot}$ km/s, which corresponds to the
lower standard curve of growth in Fig. \ref{nonstandart_cog} and agrees
with the HD and C\,I data. For a thermalized distribution
at level $l$ (open circles), the relationship between
$b_{th}$ и $b_{turb}$ should be specified. The broadening effect
will depend on this relationship. Irrespective of the relationship
between the characteristic time scales, the
case without thermalization (filled circles) can take
place for $b_{turb} \gg b_{th}$. In this case, the curve of growth
passes higher, i.e., it corresponds to larger effective $b$ 
for the standard curves of growth. Naturally, the limit
$b_{turb} \ll b_{th}$ corresponds to complete thermalization,
with the nonstandard curve of growth coinciding with
the standard one for $b=1.8 \,\mbox{km/s}$. Second, the
shape of the curve of growth depends on the column
density at the ground level. This is easy to explain,
since the shape of the distribution (Fig.~\ref{populating}, panels (c))
depends on the column density at the ground level in
the cloud via the line profiles (Fig.~\ref{populating}, panels (a)). In
the case under consideration, the column density was
chosen to be $N_{0}=10^{19}\, \mbox{cm}^{-2}$. 


It should be noted that the broadening effect manifests
itself only in a certain range of column densities
at the ground level. The distribution is not yet
broadened at low column densities ($N_{tot}<10^{15} \,\mbox{cm}^{-2}$
for typical line parameters of the Lyman or Werner
molecular hydrogen band). At high column densities
($N_{tot}>10^{20}\, \mbox{cm}^{-2}$), the broadening effect is suppressed
by the Lorentz wings (see Fig.~\ref{populating}).


Thus, the inferred effective broadening takes place
in the case of directional radiation in the cloud and
using the standard curve of growth can lead to a systematic
error in the column densities and distribution
parameters.


\section*{A SCHEME FOR CALCULATING A MOLECULAR HYDROGEN CLOUD}
\label{Cloudmodel}
\noindent
Let us apply the broadening effect that we obtained
in the previous section to molecular hydrogen.
This effect arises when considering the transfer of
directional radiation in a medium and is related to
the peculiarity of the radiation interaction with the
energy level structure. Therefore, we will consider a
cloud composed only of molecular hydrogen. Since
the radiation directionality is important for the manifestation
of the broadening effect, we will consider
a one-dimensional cloud model, i.e., one direction
in which the radiation propagates coincident with
our line of sight. In this approximation, we neglect
the background radiation compared to the directional
one, which is quite valid when a bright (in the ultraviolet)
star (or a quasar) is close to the cloud. Naturally,
in this case, the star lies behind the cloud and the line
of sight coincides with the direction of the star. It is
important to note that in this cloud model, we do not
consider such processes and the photodissociation of
molecular hydrogen and its formation on dust 
(\cite{Hollenbach1971}, \cite{Jura1975b}). A remark about the
influence of dust will be made below.

\subsection*{Multilevel Model}
	When we pass from the three-level system to the
real multilevel model of the hydrogen molecule, a
number of peculiarities arise that should be taken into
account.
	
	The first important peculiarity is that there exist
two different states of molecular hydrogen -- ortho- and
parahydrogen. In general, the direct radiative
transitions between these states are forbidden 
(\mbox{$\tau \sim 5\times10^{12}$ yr}). A much faster mixing of the ortho- and
para-states is possible through collisions with $H^+$
and $H_3^+$ (\cite{Abgrall1992}) and during the
dissociation of hydrogen followed by its formation.
However, under our model assumptions, the ortho- and
parahydrogen systems may be considered separately.
	
	The second peculiarity consists in an allowance for
the presence of many lines in the Lyman and Werner
bands through which radiative pumping proceeds.
This means that the sum over all the possible transitions
of the Lyman and Werner bands will now appear
in Eq. \eqref{P_0k} for the photoexcitation rate. To calculate
the latter, we used the wavelengths and oscillator
strengths for the Lyman and Werner band transitions
from \cite{Abgrall_L}, \cite{Abgrall_W}, respectively. The
weight factor that describes the ro-vibrational cascade
in the ground electronic state (\cite{Dalgarno1974}) should also be added. The ro-vibrational
cascade is peculiar in that it allows high-lying rotational
levels to be populated from low-lying ones
(\cite{Jura1975a}). We will assume that the ro-vibrational
cascade occurs almost instantly, i.e., on time scales
much shorter than those of the processes considered
in our model.
	
	The next remark is that the first several rotational
levels of the hydrogen molecule should be considered
jointly, i.e., we should depart from the consideration
of the ground level as a reservoir. For example, the
levels only with even rotational numbers ($J=0, 2, 4,$\,\,
etc.) are present in the parahydrogen system in the
ground electronic state. Assuming that almost all of
the molecules are at the lower vibrational level, we
should take into account the radiative pumping of
the $J=2$ level from the $J=0$ level and vice versa.
We can restrict our analysis to the first several rotational
levels, since the time scales of the spontaneous
$J \to J-2$ transitions in the lower vibrational state
decrease with increasing  $J$; this means that the upper
rotational levels can be included in the ro-vibrational
cascade. The data for the ro-vibrational cascade that
we used from \cite{Dalgarno1974} are restricted
to the first eleven rotational levels, which are
enough under typical conditions of the interstellar
medium.
	
	Yet another peculiarity is that although we disregarded
the presence of atomic hydrogen, it can be
assumed that there is a dip in the spectrum for 
$\lambda <912$ \AA \, related to the absorption of atomic hydrogen
in the ionization continuum for the radiation incident
on the cloud. This allows us to restrict our analysis
of the transitions in the Lyman and Werner bands
to the 20th and 6th vibrational levels, respectively.
As a result, we calculated the radiative transfer for
approximately 600 resonance lines of the Lyman and
Werner H$_2$ bands.
	
	\subsection*{Balance Equations}
	
	Let us write the balance equations for the system
of molecular hydrogen levels under our assumptions.
In a stationary situation,
$$
 \frac{\partial n_i(v_z, z)}{\partial t} = 0
$$
and it is convenient to write the system of equations
to determine the particle distributions at the levels in
matrix form,
\begin{equation}
	G(v_z, z)  n(v_z, z)	= 0, 
	\label{main_matrix}
\end{equation}
where $n(v_z, z)$ is the column of the particle velocity
distributions for various rotational levels; $G$ is
the matrix that describes the transitions between the
levels,
\begin{equation}
	G(v_z, z)=B(v_z, z)+A+C,
\end{equation}
where the matrix $A$ is related to spontaneous quadrupole
transitions between rotational states, the matrix
$C$ is related to collisional transitions, and the
matrix $B$ is related to radiative pumping. The specific
form of the matrix can be easily obtained by considering
the balance between the particles going from and
coming to some level $i$. For example, the components
of the matrix $B$ can be represented as
\begin{equation}
		B_{ij} = \Pi_{ij} - \delta_{ij} \sum\limits_k \Pi_{ik},
\end{equation}
where $\delta_{ij}$ is the Kronecker symbol and $\Pi_{ij}$ is the
photoexcitation rate for the transition from the $i$th
rotational level to the $j$th one and, what is important,
is a function of $(v_z, z)$ and depends on $n_i(v_z, z)$. This
is related to radiative transfer, which manifests itself
in the saturation of the radiative pumping line, and it
is this fact that provides the required broadening. The
photoexcitation rates can be written as
\begin{equation}
	\Pi_{ij}=\sum\limits_k P_{ik}(v_z, z) K_{kj},
\end{equation}
where $k$ is the set (B(C), $\nu', J')$ ($J'$ are the permitted
rotational levels). Accordingly, $P_{ik}$ is the photoexcitation
rate for the $i \to k$ transition:
$$
  P_{ik}(v_z, z)=\int\limits_{-\infty}^{+\infty} \frac{I(\nu, z)}{h\nu} \kappa_{ik}\left(\nu+\nu \frac{v_z}{c}\right)\, d\nu,
$$
and $K_{kj}$ is the probability to relax $k=$(B(C), $\nu', J') \to ($X, $\nu=0, J=i)$; for convenience, it can be separated
into the relaxation from excited electronic levels
and the ro-vibrational cascade,
\begin{equation}
	K_{kj}=\sum\limits_\nu\sum\limits_JG_{k(\nu, J)}\beta_{(\nu, J)i} \, ,
\end{equation}
where $G$ describes the relaxation from upper levels
and $\beta$ is responsible for the cascade in the lower
electronic state. We used the coefficients of the rovibrational
cascade from \cite{Dalgarno1974}.
The collisional transition rates (matrix $C$) were taken
from \cite{Abgrall1992}.

System \eqref{main_matrix} is underdetermined, since one of the
equations is not linearly independent. Therefore, we
should exclude it and expand the system by an additional
natural condition—the condition on the total
number density. Since, as we noted above, the
ortho- and parahydrogen systems may be considered
independently, the conditions on their total number
density can be written separately:
\begin{equation}
 n_0(v_z,z)+n_2(v_z,z)+  s +n_{10}(v_z,z)= n_{para}(v_z,z),
 \label{n_para}
\end{equation}
for parahydrogen and a similar condition for orthohydrogen
(with the total number density $n_{ortho}$):
$n_{para}(v_z, z)$ and $n_{ortho}(v_z, z)$ are model-dependent
quantities. We performed the calculation by assuming
that these distributions were Maxwellian and that
the integral of $n_{para}(v_z, z) + n_{ortho}(v_z, z)$ over all
velocities was equal to the total number density of
molecular hydrogen. Assuming that these systems
are in thermodynamic equilibrium, the ratio of the
para- and orthohydrogen number densities is determined
by the temperature in the cloud (typically,
$T \approx 100^o$ K).

System \eqref{main_matrix} should be solved numerically. We
consider a homogeneous medium with typical molecular
hydrogen number densities, $n_{\rm H_2} = 10 \div 1000\, \mbox{см}^{-3}$.
$N_{tot}(z)$, the column density of molecular
hydrogen to depth $z$, can serve as a measure of the
depth of penetration into the cloud.


\section*{RESULTS}
\label{results}
\noindent

\subsection*{Main Results}

\begin{figure*}[ht!]
	\centering
		\includegraphics[width=0.85\textwidth]{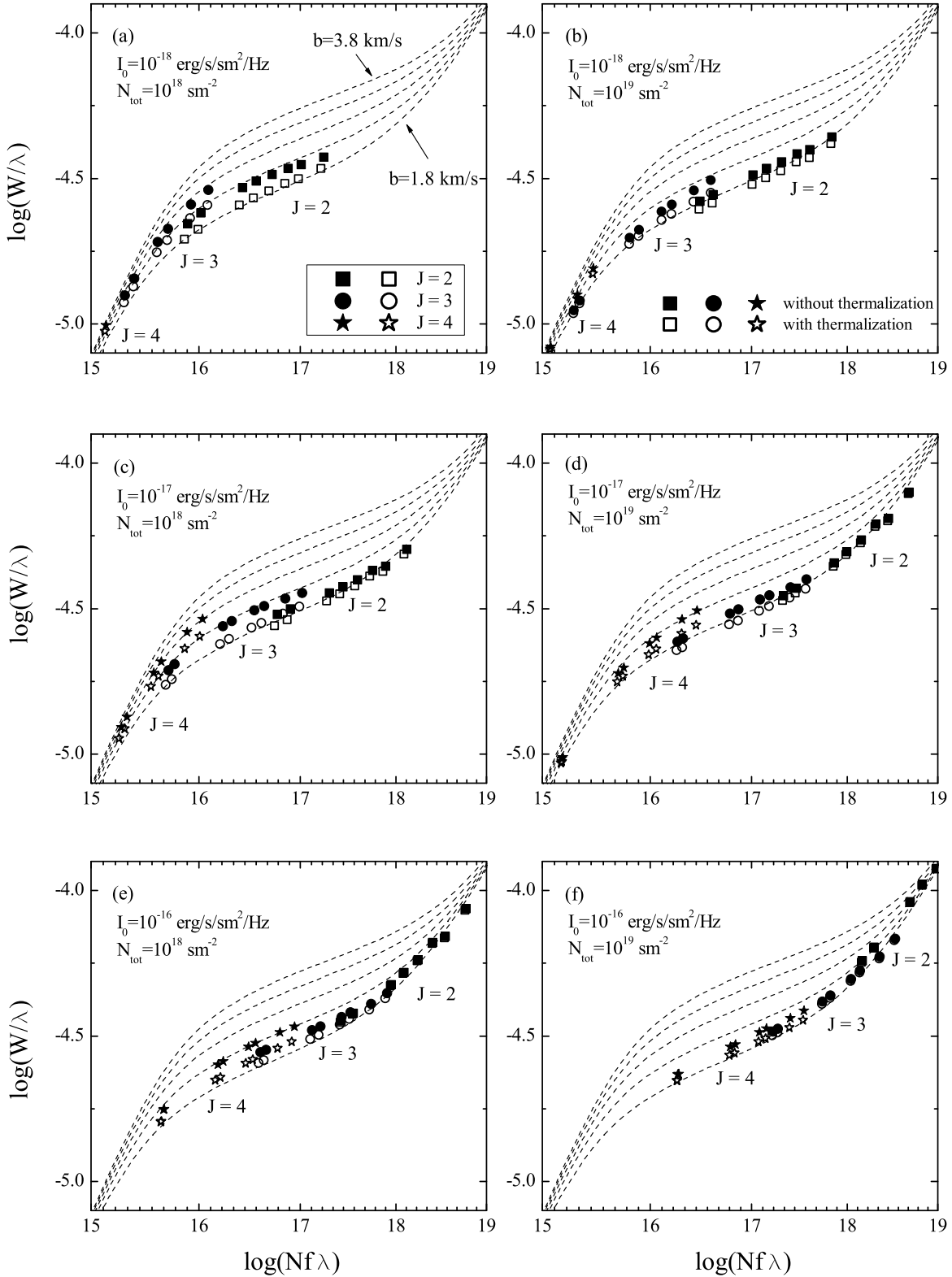}
		\caption{Results of our calculation of the molecular hydrogen cloud model for various intensities of the incident radiation $I_0$ and total molecular hydrogen column densities $N_{tot}$. The squares, circles, and asterisks indicate the equivalent widths of the lines
from the $J=2, 3, 4$ levels, respectively. The filled and open symbols represent the results of our calculation, respectively,
without and with thermalization of the velocity distribution at the level. The following distribution parameters were chosen:
$b_{th} = 0.8\,\mbox{km/s}$, $b_{turb} = 1.6\,\mbox{km/s}$. The dashed lines represent the standard curves of growth with parameters $b$ from 1.8 to $3.8\,\mbox{km/s}$ at steps of $0.5\,\mbox{km/s}$. In the absence of a broadening effect, all symbols would fall on the lower standard curve
of growth with $b = 1.8\,\mbox{km/s}$.}
 		\label{results_cogs}
\end{figure*}

The results of our calculations based on the
scheme presented in the previous section are shown
in Fig. \ref{results_cogs} for various combinations of $N_{tot}$ (the
total molecular hydrogen column density in the
cloud) and $I_{0}$ (the intensity of the incident directional
radiation). For two column densities $N_{tot}$,
$10^{18}\,\mbox{cm}^{-2}$ и $10^{19}\, \mbox{cm}^{-2}$, we chose three different intensities
of the incident radiation: $10^{-16}$, $10^{-17}$ и $10^{-18}$ $\mbox{erg/cm}^2\mbox{/s/Hz}$. 
Thus, there are a total of
six different calculations (Fig. \ref{results_cogs}). The background
isotropic radiation was assumed to be negligible
compared to the directional one (this is valid when
the directional radiation is determined by a nearby
star or a quasar). For all calculations, the distribution
parameters were chosen from observational
data on HD and C\,I in the \qso\,\, absorption system: $b_{th} = 0.81\, \mbox{km/s}$ for the thermal
distribution and $b_{turb} = 1.6\, \mbox{km/s}$ for the turbulent
one; as a result, the Doppler parameter $b$ for the
distributions $n_{para}(v_z)$ and $n_{orto}(v_z)$ (see \eqref{n_para}) is
$b=\sqrt{b_{th}^2+b_{turb}^2} = 1.8 \, \mbox{km/s}$. The total molecular
hydrogen number density was assumed to be uniform
and the calculation was performed for several values
of $n_0=n_{para}+n_{orto} = 10, 100, 1000 \,\,\mbox{cm}^{-3}$.


As the result of our calculation, Fig. \ref{results_cogs} presents the
dependences of the line profiles for the Lyman and
Werner molecular hydrogen bands on the depth of
radiation penetration into the cloud. The line profiles
reflect the particle velocity distribution on the line
of sight at these levels and are calculated consistently.
To present our results, we chose several lines
for which the equivalent width was calculated and
plotted on the curve of growth. For comparison with
the observational data for the absorption system of
\qso\,\,\,, we used the transitions from the
rotational $J=2$(squares), $J=3$(circles), and $J=4$
(asterisks) levels. Figure \ref{results_cogs} present two extreme cases:
the absence of thermalization (filled symbols) and
complete thermalization at all levels (open symbols)
for the chosen parameters $b_{th}$ и $b_{turb}$.


As was shown above, one might expect a significant
broadening of the particle velocity distribution
(by almost a factor of 2 in terms of the effective
Doppler parameter) in the three-level system.
However, the broadening of the distribution in the
cloud model decreases appreciably. The main reason
is that there are many radiative population transitions.
For the chosen rotational level, there are about 60
such transitions in the Lyman or Werner bands. Each
of these transitions has its own values of $\lambda f$ — the
product of the transition wavelength by the oscillator
strength. For each of them, the broadening effect will
begin (with line saturation) and be suppressed (when
the Lorentz wings appear) at different depths in the
cloud. This is easy to understand, since the optical
depth includes the combination $N \lambda f$. Therefore,
the incompletely broadened distributions from some
transitions and the distributions with the influence
of the Lorentz wings (reducing the broadening) from
other transitions will be admixed to the maximum
broadening of the distribution from some transition at
some depth.


In addition to the aforesaid, the situation will be
complicated at a high intensity of the incident radiation.
In this case, the excited rotational levels will be
populated more heavily than the ground levels until
the lines of various transitions become saturated and
the photoexcitation rate decreases. This will be true
for the cloud shell. We will consider the peculiarity
that arises in this case below.


When the curve of growth is analyzed, it is necessary
that the lines from the levels with a broadening
fall on the logarithmic segment of the curve of growth.
Otherwise, if the broadening is present, we will not
see it on the linear and square-root segments due to
the peculiarity of the curve of growth. Therefore, when
the system of molecular hydrogen is considered, the
dependence in the manifestation of the broadening
effect on the intensity of the radiation incident on
the cloud is obvious. The excited rotational levels will
be populated more heavily with increasing intensity
of the incident radiation. This is clearly seen if we
compare panel (a) in Fig. \ref{results_cogs} with panels (c) and (e)
(and panel (b) with panels (d) and (f)). In this case, the
points corresponding to the transitions from different
rotational levels become closer to one another (cf. the
squares, circles, and asterisks).


The broadening effect in the chosen six combinations
of parameters reaches its maximum at
$N_{tot}= 10^{18}\, \mbox{cm}^{-2}$, $I_0 =10^{-17}\, \mbox{erg/cm}^2\mbox{/s/Hz}$
(panel (c) in Fig. \ref{results_cogs}). This can be explained as follows:
the total hydrogen column density $N_{tot}$ should not
be high ($>10^{19}\, \mbox{см}^{-2}$); otherwise, the Lorentz wings
begin to suppress the effect (see Figs. \ref{results_cogs}b, \ref{results_cogs}d, and
\ref{results_cogs}f). Obviously, the radiation must be intense enough
for the excited rotational levels to fall into the region
where the logarithmic part of the curve of growth
begins, where the effect is most pronounced. If the intensity
is higher, then the excited levels are populated
more heavily (Fig. \ref{results_cogs}e) and approach the square-root
part of the curve of growth; if the intensity is lower
(Fig. \ref{results_cogs}a), then the levels fall on the linear part of the
curve of growth. On these segments of the curve of
growth, the broadening effect does not manifest itself,
because it is insensitive to $b$.


However, the transfer of directional radiation that
we considered is not enough to explain the broadening
of the distribution for excited rotational levels
of molecular hydrogen that arises in the absorption
system of \qso. Therefore, additional possibilities
for the explanation should be sought. Nevertheless,
this effect cannot be neglected, provided that
the radiation directionality in the cloud is dominant.


\subsection*{The Influence of Dust}

One of the ways to explain the broadening of the
velocity distribution is the formation of molecular hydrogen
on dust (\cite{Lacour2005}). However, the
dust-related broadening effect during the propagation
of directional radiation can be only a small correction.
When a stationary situation is considered, the
amount of formed hydrogen is assumed to be exactly
equal to the amount of dissociated one, which accounts
for 13\% of the molecules excited in the Lyman
or Werner bands. The remaining 87\% are involved in
radiative pumping, which, as was shown above, leads
to a broadening of the distribution. This means that
the two effects will be added with weights of 0.87
and 0.13 for the broadenings caused by radiative
pumping and dust formation, respectively, relative to
the distribution functions.


The presence of dust in the cloud can lead to a
slightly different effect. The point is that dust reduces
the intensity in the continuum. In fact, the population
by the Lorentz wings also takes place in the continuum.
Therefore, in the presence of dust in the cloud, it
is capable of eating away the ultraviolet continuum in
such a way that starting from some hydrogen column
density $N_{d}$ (depending on the dust number density
in the cloud), it essentially will reduce the population
in the Lorentz wings to zero, i.e., in an ideal case, it
will only “freeze” the broadening. We considered this
additional effect, but it will not allow us to obtain the
additional broadening needed to explain the observational
data.



\subsection*{The Model of a ”Hot Shell”}

\begin{figure*}[t!]
	\centering
		\includegraphics[width=0.68\textwidth]{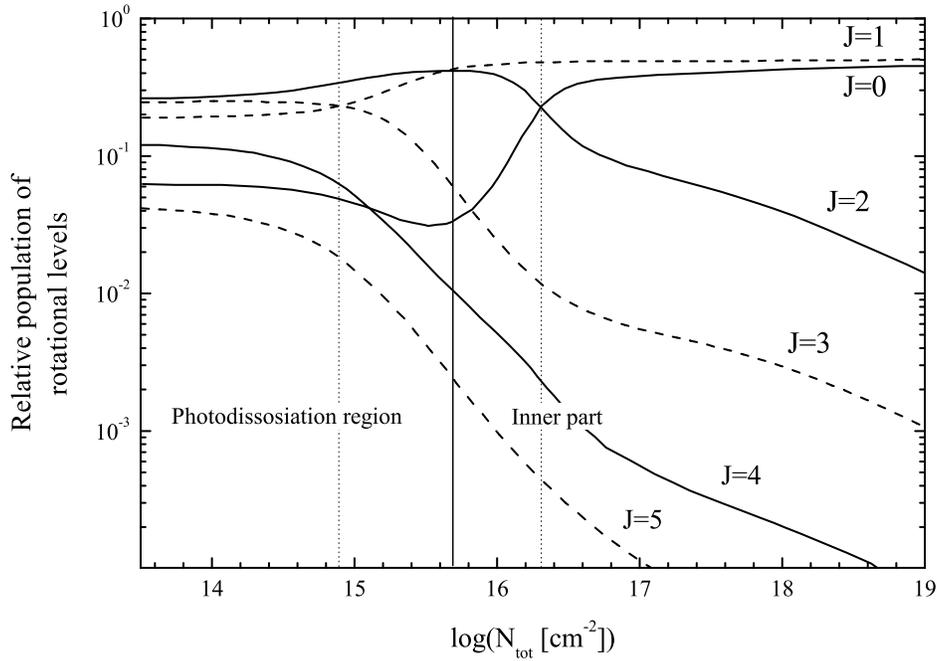}
		\caption{Populations of various rotational levels versus depth of penetration into the cloud. The cloud is arbitrarily separated by
the vertical solid line into the shell (photo-dissociation region) and the inner part. The dotted vertical lines mark the depths at
which the $J=1$ and $J=3$ (orthohydrogen) populations and the $J=0$ and $J=2$ (parahydrogen) populations become equal.}
 		\label{results_conc}
\end{figure*}

The assertion of \cite{Noterdaeme2007} that
the velocity distribution at excited rotational levels
can be broadened in the presence of a photodissociation
region seems the most realistic explanation. This
explanation naturally also arose in our paper when
we considered the results of our model calculation
presented in Fig. \ref{results_conc}.


In Fig. \ref{results_conc}, the populations of various rotational
levels are plotted against the depth of radiation penetration
into the cloud. We see from the figure that the
cloud can be arbitrarily separated into two regions:
the shell and the inner part. In the shell, the lines of the
Lyman and Werner bands are not saturated and the
radiative pumping rate is high at a sufficiently high intensity
(in our case, \mbox{$I_0=10^{-17}\, \mbox{ers/cm}^2\mbox{/s/Hz}$}).
Consequently, the transition rates between the rotational
levels of molecular hydrogen in the lower
vibrational (ground) electronic state are also high. In
that case, the excited rotational levels will be populated
significantly (the $J=2$, $3$ and $4$ levels are
populated more heavily than the ground ones). Since
the absorption lines become saturated as the radiation
penetrates into the cloud (starting from $N_{tot} \approx 10^{15}\, \mbox{cm}^{-2}$), the photoexcitation rates and, hence,
the population of the excited rotational levels decrease.
As a result, only the ground levels turn out
to be populated inside the cloud (in our case, $N_{tot} \gtrsim 10^{16} \, \mbox{cm}^{-2}$).We also see that the inner part of a cloud
with $N_{tot}>10^{18}\, \mbox{cm}^{-2} $ (measured in the molecular
hydrogen column density) will be several orders of
magnitude larger than its shell. Therefore, the ground
levels manifest themselves and allow us to determine
the physical conditions only in the inner part of the
cloud. As regards the excited levels, the column density
of the excited rotational levels in the shell and the
inner part of the cloud may turn out to be comparable
in view of the radiative population pattern. This
means that the excited rotational levels will reflect the
physical conditions in the shell.

Since the photoexcitation rates in the shell are
higher, the photodissociation rates are also higher.
Photodissociation can heat the medium, because an
energy of $4,48$ eV is released in one act of this process.
Therefore, in the regions with active photodissociation,
the medium can heat up and, as a result,
have a broader velocity distribution of the molecules.
These are called photodissociation regions (PDRs)
(see, e.g., \cite{Hollenbach1999}, \cite{LePetit2006}).

\begin{figure*}
	\centering
		\includegraphics[width=0.65\textwidth]{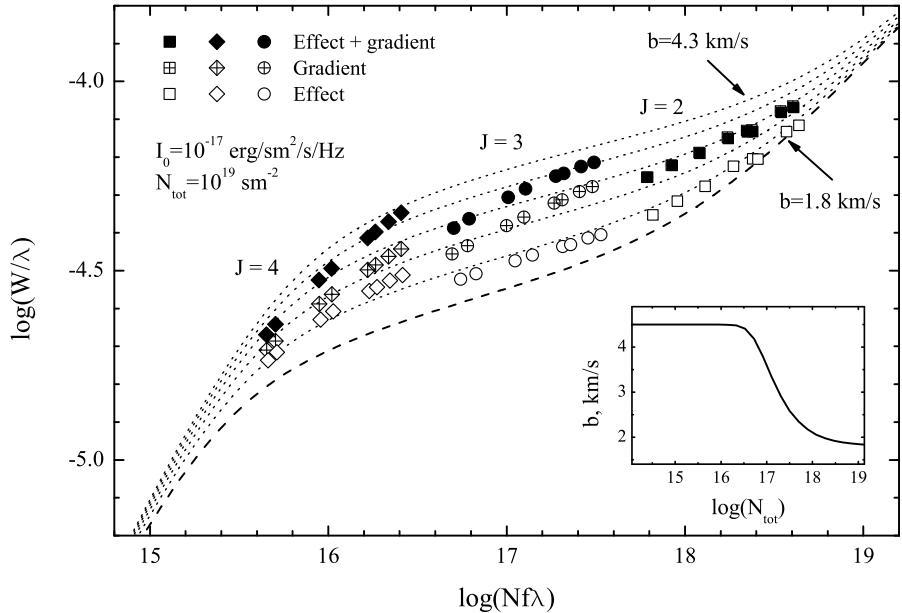}
		\caption{Results of our calculation of the molecular cloud model in the presence of a hot shell introduced through a nonuniformity
of $b$ over the cloud (shown in the insert). The filled, open, and crossed symbols indicate the results of our calculation,
respectively, without and with the shell and with a nonuniformity of $b$ but without any nonequilibrium radiative population.
The dashed lines indicate the standard curves of growth with parameters b from 1.8 to $3.8\, \mbox{km/s}$ at steps of $0.5\, \mbox{km/s}$.}
 		\label{results_pdr}
\end{figure*}

We will take into account the PDR by specifying
a nonuniformity of the Doppler parameter $b$ in the
cloud. The medium in the shell will have a higher
value of $b$ than that inside the cloud. In general, $b$ can
increase in the shell not only through an increase in
$b_{th}$, i.e., the heating of the medium, but also through
an increase in the turbulent velocity component $b_{turb}$.
To specify the gradient in $b$, we will take into account
the fact that the photodissociation rate (and, hence,
the heating of the medium) is related to the photoexcitation
rate. The dependence of $b$ on the depth of
penetration into the cloud that we chose is shown in
Fig. \ref{results_pdr}. We chose the characteristic depth at which
the shell–inner part transition occurs to be $N_{tot}=3 \times 10^{16}\, \mbox{cm}^{-2}$. 

\begin{figure*}
	\centering
		\includegraphics[width=0.55\textwidth]{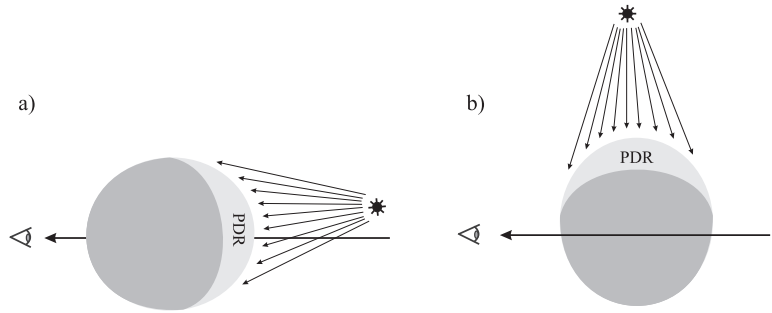}
		\caption{Possible geometric configurations of PDRs relative to the line of sight: (a) the line of sight is directed toward the source (a young star, a quasar) and, therefore, a PDR can be observed; in this case, the broadening effect should be taken into account;
(b) the line of sight is perpendicular to the direction of the star and the manifestation of a PDR in observations is unlikely; in
this case, the broadening effect is suppressed.}
 		\label{pdr_geometry}
\end{figure*}

Figure \ref{results_pdr} presents the results of our calculation of
the model for radiative transfer in a molecular hydrogen
cloud with a PDR specified by a nonuniformity
of the parameter $b$. The open symbols indicate the
equivalent widths of the lines from various rotational
H$_2$ levels obtained without including the PDR but
in the presence of a nonequilibrium radiative population
(this calculation was presented above, in Fig. \ref{results_cogs},
panel (d)). The filled symbols represent the results of
our calculation when adding a $b$ nonuniformity shown
in the insert in Fig. \ref{results_pdr}. We see that the line equivalent
widths for the transitions from excited rotational
levels in the presence of a $b$ nonuniformity lie high,
near the curves of growth with $b \sim 4.0\,\mbox{km/s}$. This
result shows that the presence of a PDR can be used
to explain the observational data for the absorption
system of \qso. Because of the simplified
scheme of our calculation, we did not attempt to
determine the exact parameters of the nonuniformity
but only show the importance of taking into account
the PDR.

 
 We also see from Fig. \ref{results_pdr} that when applying an
analysis of the standard curve of growth to the computed
points in the presence of a PDR (filled symbols),
to achieve correspondence to some standard
curve of growth, it will be necessary to displace the
points greatly leftward. This will lead to an even larger
increase in the effective parameter $b$ and larger errors
in the populations of excited rotational levels.
Therefore, it is important to note that assigning the
same Doppler parameter to all rotational levels of
molecular hydrogen when analyzing the absorption
systems can lead to serious systematic errors in the
column density and other physical parameters.


	The crossed symbols indicate the results of our
calculation in the absence of any nonequilibrium radiative
population, i.e., the velocity distributions at
all rotational levels were assumed to correspond to
a Maxwellian distribution with a nonuniform $b$ over
the cloud. We see that the effect should be taken
into account for a detailed analysis of PDRs. If we
associate the presence of a PDR with the location of
the cloud near a young bright star, then this effect
should be taken into account, because, in this case,
the radiation can be assumed to be directional, i.e., to
propagate near the line of sight. The point is that a
PDR close to a star must be located in the cloud on
the side of the star (see Fig. \ref{pdr_geometry}a). In the opposite case,
where the “young star–cloud” line is perpendicular
to the line of sight, the effect of a nonequilibrium radiative
population will be unobservable on the cloud.
However, in this geometry, the PDR is also difficult
to observe, because the molecular hydrogen column
density is low ($N_{0} < 10^{15}\, \mbox{cm}^{-2}$) in this region (see
Fig. \ref{pdr_geometry}b). In this case, if the line of sight passes
through the PDR, then it will not pass through the
cloud interior, where the bulk of the molecular hydrogen
is concentrated, and, conversely, if the line
of sight passes through the interior, it will not touch
the PDR.

\section*{CONCLUSIONS}
\noindent
\label{conclusion}

Observations of molecular hydrogen clouds reveal
a broadening of the velocity distribution at excited
rotational levels. This effect is present in the observational
data for Galactic (\cite{Jenkins1997}, \cite{Lacour2005}) and 
high-redshift (\cite{Noterdaeme2007}) clouds. We suggested a mechanism that
could be responsible for the effective broadening of
the distribution at excited rotational levels. The condition
for the manifestation of this effect is radiation
directionality. Using a three-level model system, we
showed that the radiative pumping of an excited level
for directional radiation could lead to a broadening
of the particle velocity distribution at this level. The
broadening of the distribution was considered in two
limits: with thermalization of the velocity distribution
(the turbulent component is retained) and without
its thermalization. In the absence of thermalization,
our mechanism in the three-level system considered
(at transition line parameters typical of molecular
hydrogen) can lead to an almost twofold effective
broadening of the distribution in an analysis using the
curve-of-growth method. The suppression of the effect
during thermalization depends on the ratio of $b_{turb}$ 
and $b_{th}$, the parameters of the Maxwellian distribution
attributed to turbulence and thermal motions

We applied the inferred effect to the system of
molecular hydrogen energy levels and calculated a
simple molecular hydrogen cloud model. We considered
the balance at the excited rotational levels
of the lower vibrational ground electronic state by
taking into account the set of lines of the Lyman and
Werner bands through which the radiative pumping
proceeds. We also took into account the presence of a
ro-vibrational cascade in the ground electronic state.
The large number of lines slightly reduces the effect,
because its manifestation in the radiative level population
is heterogeneous. However, the broadening of
the velocity distribution of molecules during radiative
pumping is not enough to explain the data on the
broadening of the distribution at excited rotational
levels obtained by analyzing the absorption system of
\qso.
Based on the calculated populations of rotational
levels with depth, we assumed an additional mechanism
capable of producing an effective broadening.
This mechanism is related to the possibility of describing
a molecular cloud by two different regions:
the shell and the inner part. This separation, which
naturally arises when the radiative transfer is considered
in the cloud, stems from the fact that the
excited and ground rotational levels are mostly populated
in the shell and the inner part, respectively.
In that case, the shell will be a PDR in which the
influence of ultraviolet photons on the chemical composition
and thermodynamic balance of the medium
is strong. In view of the assumptions made in our
model, we introduced a spatial nonuniformity of the
Doppler parameter $b$ to take into account the PDR.
The increase in $b$ in the shell is related to the heating
of the medium or to the stronger turbulence. With a
nonuniform parameter $b$, we can obtain the broadenings
of the distributions at excited rotational levels
needed to explain the observational data. A similar
broadening mechanism for the velocity distributions
of molecules was used by \cite{Noterdaeme2007}.
However, they did not consider the effect that we
showed and that is related to radiation directionality.
We point out that when a PDR manifests itself in the
observations, the radiation directionality effect should
also manifest itself. Therefore, in the case of a PDR
region, it should be taken into account to eliminate
the systematic error when the observational data are
analyzed.

\section{ACKNOWLEDGMENTS}
\noindent

This work was supported by the Russian Foundation
for Basic Research (project no. 08-02-01246-a)
and the Program of the Russian President for Support
of Scientific Schools (project no. NSh-2600.2008.2).

\section*{APPENDIX}

To derive the distribution thermalization while the
turbulent component is retained, we will assume that
that the distribution at the ground level in \eqref{n_l}
is an equilibrium one and use the assumption about
microturbulence. We can then write
\begin{equation}
   n_l(v_z, z) = \frac{n_0(z)}{A_l} P_{0k}(v_z, z) \int\limits_{-\infty}^{+\infty}dv_t \, w_{turb}(v_t) w_{th}(v_z-v_t),
   \label{n_therm}
\end{equation}
where $n_0(z)$ is the number density at the ground
level and $P_{0k}$ is defined by Eq. \eqref{P_0k}; $w_{turb}$ and $w_{th}$ are
the Gaussian functions with parameters $b_{turb}$ and $b_{th}$ ,
respectively. Here, we used the fact that the particle
velocity distribution was a convolution of the thermal
and turbulent distributions.

Integrating Eq. \eqref{n_therm} over $v_z$ yields
\begin{equation}
   n_l(z)= \frac{n_0(z)}{A_l}\int\limits_{-\infty}^{+\infty}dv_t \int\limits_{-\infty}^{+\infty}dv_z \, w_{turb}(v_t) w_{th}(v_z-v_t) P_{0k}(v_z, z).
   \label{n_therm2}
\end{equation}
We thus see that the distribution of the upward thrown
particles in turbulent velocity is
\begin{multline}
   w'_{turb}(v_t, z)=\frac{n_0(z)}{n_l(z)}   \frac{1}{A_l}   w_{turb}(v_t) \times \\ 
   \times \int\limits_{-\infty}^{+\infty}dv_z \, w_{th}(v_z-v_t) P_{0k}(v_z, z).
   \label{turb}
\end{multline}

Under our assumptions, this nonequilibrium turbulent
distribution is retained. Thus, after the thermalization
of the medium, the particle distribution at
level $l$ can bewritten as
\begin{equation}
	 \tilde n_l(v_z, z)= n_l(z)  \int\limits_{-\infty}^{+\infty}dv_t \, w'_{turb}(v_t, z) w_{th}(v_z-v_t).
\end{equation}
Substituting \eqref{turb} here, we obtain
\begin{multline}
	 \tilde n_l(v_z, z)= \frac{n_0(z)}{A_l}   \int\limits_{-\infty}^{+\infty}dv'_z \, P_{0k}(v'_z, z) \times \\ 
	 \times \int\limits_{-\infty}^{+\infty} dv_t \,w_{th}(v'_z-v_t) w_{turb}(v_t) w_{th}(v_z-v_t).
	 \label{n_tilde}
\end{multline}
Integration over $v_t$ yields the product of twoGaussian
functions with parameters $b$ and $b_{+}=\sqrt{b^2_0+b^2_{th}}=b_{th}\sqrt{1+(b_{turb}/b)^2}$.

Equations \eqref{n_tilde} and \eqref{n_therm2} then give \eqref{thermalize}.


\begin{thebibliography}{}
		\bibitem[Abgrall et al. 1992]{Abgrall1992} Abgrall H., Le Bourlot J., Pineau Des Forets G., et al., Astron. Astrophys. \textbf{253}, 525, (1992).
		\bibitem[Abgrall et al. 1993a]{Abgrall_L} Abgrall H., Roueff E., Launay F., et al., Astron. Astrophys. Suppl. Ser. \textbf{101}, 273, (1993а).
		\bibitem[Abgrall et al. 1993b]{Abgrall_W} Abgrall H., Roueff E., Launay F., et al., Astron. Astrophys. Suppl. Ser. \textbf{101}, 323, (1993б).
		\bibitem[Bluhm et al. 2003]{Bluhm2003} Bluhm H., de Boer K. S., Marggraf O., et al., Astron. Astrophys. \textbf{398}, 983 (2003)
		\bibitem[Varshalovich et al. 2001]{Varshalovich2001} Varshalovich D.A., Ivanchik A.V., Petitjean  P., Astron., letters \textbf{27}, 803 (2001)
		\bibitem[Dalgarno \& Black 1974]{Dalgarno1974} Dalgarno A. \& Black J.H., Astrophys.J. \textbf{203}, 132, (1974).
		\bibitem[Jenkins \& Peimbert 1997]{Jenkins1997} Jenkins E. B. \& Peimbert A., Astrophys. J. \textbf{477}, 265, (1997).
		\bibitem[Ge \& Bechtold 1999]{Ge1999} Ge J. \& Bechtold J., APS confirence series \textbf{156}, L121, (1999)
		\bibitem[Draine 1978]{Draine1978} Draine B.T., Astrophys.J. Suppl. Ser. \textbf{36}, 595,  (1978)
		\bibitem[Ivanchik et al. 2005]{Ivanchik2005} Ivanchik A., Petitjean P., Varshalovich D., et al. Astron. Astrophys. \textbf{440}, 45, (2005)
		\bibitem[Lacour et al. 2005]{Lacour2005} Lacour S., Ziskin V., Hebrard G., et al., Astrophys. J. \textbf{627}, 251, (2005).
		\bibitem[Le Petit et al. 2006]{LePetit2006} Le Petit F., Nehme C., Le Bourlot J., et al., Astrophys. J. Suppl. Ser. \textbf{164}, 506, (2006).
		\bibitem[Levshakov \& Varshalovich 1985]{Levshakov1985} Levshakov S.A. \& Varshalovich D.A., MNRAS \textbf{212}, 517 (1985)
		\bibitem[Ledoux et al. 2003]{Ledoux2003} Ledoux C., Petitjean P., Srianand R., et al., MNRAS \textbf{346}, 209, (2003)
		\bibitem[Noterdaeme et al. 2007]{Noterdaeme2007} Noterdaeme P., Ledoux C., Petitjean P., et al., Astron. Astrophys. \textbf{474}, 393, (2007).
		\bibitem[Noterdaeme et al. 2008]{Noterdaeme2008} Noterdaeme P., Ledoux C., Petitjean P., et al., Astron. Astrophys. \textbf{481}, 327, (2008).
		\bibitem[Petitjean et al. 2000]{Petitjean2000} Petitjean P., Srianand R., Ledoux C., et al., Astron. Astrophys., \textbf{364}, L26–L30 (2000)
		\bibitem[Rachford et al. 2002]{Rachford2002} Rachford B.L., Snow T. P., Tumlinson J., et al., Astrophys.J. \textbf{577}, 221 (2002)
		\bibitem[Spitzer \& Cochran 1973]{Spitzer1973} Spitzer  L. J. \& Cochran W. D., Astropsyh.J. \textbf{186}, 23, (1973).
		\bibitem[Stecher \& Williams 1967]{Stecher1967} Stecher T.P. \& Williams D.A., Astrophys.J. \textbf{149}, 29, (1967).
		\bibitem[Savage et al. 1977]{Savage1977} Savage B.D., Bohlin R. C., Drake J. F., et al., Astrophys.J. \textbf{216}, 291 (1977)
		\bibitem[Tumlinson et al. 2002]{Tumlinson2002} Tumlinson J., Shull J. M., Rachford B. L., et al., Astrophys.J. \textbf{556}, 857 (2002)
		\bibitem[Hollenbach et al. 1971]{Hollenbach1971} Hollenbach D.J., Werner M.W., Salpeter E.E., Astrophys.J. \textbf{163}, 155, (1971).
		\bibitem[Hollenbach \& Tielens 1999]{Hollenbach1999} Hollenbach D.J., Tielens A.G.G.M., Rev. Mod. Phys., \textbf{71}, 1, (1999)
		\bibitem[Shull \& Beckwith 1982]{Shull1982} Shull J.M. \& Beckwith S., Ann. Rev. Astron. Astrophys. \textbf{20}, 163, (1982) 
		\bibitem[Shull et al. 2000]{Shull2000} Shull J.M., Tumlinson J., Jenkins E. B., et al., Astrophys.J. \textbf{538}, 73 (2000)
		\bibitem[Srianand et al. 2008]{Srianand2008} Srianand R., Noterdaeme P., Ledoux C., et al., Astron. Astrophys. \textbf{482}, L39 (2008)
		\bibitem[Srianand et al. 2005]{Srianand2005} Srianand R., Petitjean P., Ledoux C., et al., MNRAS \textbf{362}, 549, (2005)
		\bibitem[Jura 1975а]{Jura1975a} Jura M., Astrophys. J. \textbf{197}, 575, (1975а).
		\bibitem[Jura 1975б]{Jura1975b} Jura M., Astrophys. J. \textbf{197}, 581, (1975б).
\end{thebibliography}
\end{document}